# Hydrogen plasma-assisted atomic layer epitaxy of superconducting titanium nitride

Yi-Hsun Chen[1*], Yin-Chun Huang[2], Zachary Degnan[1], David Sommers[1], Kaijian Xing[3,4], Manjith Bose[5], Eduardo Solano[6], David Cortie[7], Michael Fuhrer[3], Julian A. Steele[1], Peter Jacobson[1], Miin-Jang Chen[2,8*], Arkady Fedorov[1]

1 School of Mathematics and Physics, University of Queensland, St. Lucia, Queensland 4072, Australia
2 Graduate School of Advanced Technology, National Taiwan University, Taipei 10617, Taiwan
3 School of Physics & Astronomy, Monash University, Clayton, VIC, Australia
4 Shanghai Key Laboratory of High Temperature Superconductors, Department of Physics, Shanghai University, Shanghai 200444, China
5 School of Physics, The University of Melbourne, Parkville, 3010, VIC, Australia
6 CD-SWEET beamline ALBA Synchrotron Light Source, Cerdanyola del Vallès, Spain
7 Australian Centre for Neutron Scattering, Australian Nuclear Science and Technology Organisation, Sydney, New South Wales, Australia
8 Department of Materials Science and Engineering, National Taiwan University, Taipei, 10617, Taiwan

Corresponding email: yihsun.chen@uq.edu.au

## ABSTRACT

Atomic layer deposition (ALD) offers precise, conformal control of thin-film growth and is a workhorse for semiconductor manufacturing, but its use in superconducting quantum circuits is constrained by the need to simultaneously control crystallinity, stoichiometry and interfacial disorder at cryogenic temperatures. Here we develop hydrogen plasma assisted ALD to epitaxially grow superconducting titanium nitride (TiN) thin films, using a hydrogen plasma step to modify the surface chemistry to suppress precursor-derived impurities during each deposition cycle. Synchrotron X-ray scattering reveals semi-coherent epitaxy of TiN on c-plane sapphire, with discrete crystallographic domains and minimal long-range structural disorder. Complementary X-ray spectroscopy and neutron reflectometry show negligible oxygen-related disorder throughout the film and no substantial hydrogen incorporation. The resulting TiN films exhibit a superconducting transition at 2.2 K, a kinetic inductance of 15 pH/sq and a superconducting coherence length of 14.6 nm, comparable to the measured crystal coherence length. These results establish hydrogen plasma assisted ALD as a route to structurally ordered superconducting TiN thin film epitaxy and demonstrate that the nanoscale control of ALD can be extended to materials suitable for cryogenic quantum technologies.

## INTRODUCTION

Atomic layer deposition (ALD) gained prominence in the semiconductor industry as a solution to the minimum linewidth and high aspect ratio features encountered in modern chip designs. The aggressive pursuit enables precise and conformal deposition of metals, dielectrics, semiconductors, and even superconductors.[1–3] This capability has attracted significant interest in the quantum device research community as superconducting devices are moving out of the research lab and into manufacturing of quantum computers.[4–6] The precise control of growth mechanism offers advantages of conformal coatings over complex patterned surfaces, high throughput, and scalability, which could also facilitate the future wafer-scale production of quantum devices and reduce material imperfections associated with major internal loss mechanisms such as two-level system (TLS) limiting qubit coherence in superconducting circuits.[7–9]

Elemental superconductors remain the most widely used materials in superconducting quantum circuits[10–15], but ALD grown nitrides are increasingly attractive for all-nitride device architectures[16], vertical Josephson junction fabrication[16,17], superconducting qubits[18,19], and high-kinetic inductance applications[20,21], driving the demand for the adoption of new deposition techniques. ALD techniques enable the layer-by-layer growth at the nanometre-scale through the self-limiting reaction at a relatively low temperature. However, limited surface reaction kinetics reduce adatom mobility and suppress surface diffusion which result in incomplete removal of precursor ligands leaving carbon, oxygen, and precursor-related residues throughout the film[22–24]. Plasma-enhanced ALD can achieve a better stoichiometry and crystallinity compared with conventional ALD due to high-energy reactive plasma[25,26], whereas reducing carbon- or oxygen-related impurities in metal nitrides remains a challenge due to the precursor-related residues.[27–29] Since impurities could contribute to dielectric loss associated with TLS loss which limits the performance of superconducting circuits[30–32], the development of high-precision thin film deposition methods, particularly ones that are scalable and compatible with existing industrial processes, will directly impact the success of TiN-based superconductors and quantum technologies.

Here, we introduce a hydrogen plasma step into each ALD cycle, via so-called hydrogen plasma assisted ALD (HPA-ALD). This approach overcomes the issues of impurity incorporation during regular ALD by $H_2/N_2$ mixed plamsa[33]. However, it remains unclear whether this method can be translated to grow superconductors for quantum device applications. In this work, we advance the HPA-ALD technique and demonstrate semi-coherent epitaxy of high-quality superconducting TiN thin films. Synchrotron-based characterisations reveal that our TiN exhibits a highly coherent crystal structure with minimal interface disorder or chemical impurities. Electrical transport measurements show a superconducting transition temperature ($T_c$) at 2.2 K and a corresponding kinetic inductance of 15 pH/sq. Moreover, we obtained a superconducting coherence length of 14.6 nm, which is comparable to the crystal coherent length. By combining advanced synchrotron characterisation and cryogenic transport measurements, our findings extend the functionality of HPA-ALD to superconducting materials, paving the way to realise high-quality nitride-based superconducting devices.

## RESULTS AND DISCUSSION

TiN thin films were epitaxially grown on c-plane sapphire at a process temperature of 300°C. In each ALD cycle (see Figure S1), titanium precursor was first introduced to enable chemisorption on the substrate surface, followed by an argon purge to remove excess precursor molecules and residual byproducts. Owing to the self-limiting surface reactions inherent to ALD, precise control over film thickness and growth behaviour can be achieved with nm-resolution. We then sequentially introduced $H_2$ and $N_2$ plasma separated by an additional step of Ar purging. $H_2$ plasma was first applied as a surface pre-treatment step to regulate the surface chemical state with the purpose of removing residual organic ligands and suppress carbon-related impurities[33,34]. After purging the sample with Ar, $N_2$ plasma was introduced to supply reactive nitrogen species, enabling the formation of Ti–N bonds and completing the film growth. A detailed description of the epitaxial process is provided in the Method. The sequential plasma design enhances the overall surface reaction efficiency and is consistent with the high crystallinity observed in structural analyses discussed below.

Crystallographic orientation and epitaxial alignment of TiN were characterised using synchrotron-based grazing incidence wide-angle x-ray scattering (GIWAXS). Figures 1a–c shows a representative overview of the GIWAXS data recoded as a function of the rotation angle normal to sample surface, φ, and corresponding integrated profiles. In the full 1D integrated profile (Figure 1a), Bragg peaks correspond to the (111), (200), and (220) planes of space group *Fm-3m*. The unit cell size is estimated to be 4.197 Å based on a structural refinement (using Le Bail method), which is ~1% smaller compared to that of bulk TiN (4.241 Å)[35]. Examination of the 1D profiles recorded during φ-rotation shows that TiN exhibits Bragg (220) peaks with a period of 60° (Figure 1b), indicating a twinned structure[33]. 2D GIWAXS images allow us to investigate crystalline texture in both the out-of-plane and in-plane directions. Figure 1c shows a 2D GIWAXS image of the TiN film obtained at an incident angle $\alpha = 0.28°$, which is above the critical scattering angle and probes the entire thickness of the TiN film (Figure S2). The Bragg peaks of the twinned structure appear as discrete spots without an azimuthal dispersion, indicating a highly coherent epitaxial structure[36]. Since the measurement probes the bulk-sensitive scattering regime, these discrete Bragg peaks also indicate the absence of long-range disorder throughout the film or more complex mesostructure, including at the interface between TiN and sapphire. The high degree of structural order allows us to reconstruct the crystal structure of TiN films in real space. In the 2D GIWAXS reciprocal space image, the azimuthal angle of a Bragg peak represents the orientation of a specific crystal plane with respect to the substrate. The (111), (200), and (220) planes at an azimuthal angle of 21.1°, 38.5°, and 59.1° respectively gives a cube-on-hexagon orientation relationship TiN $[1\ 1\ \bar{2}]$ ∥ $Al_2O_3$ $[2\ \bar{1}\ \bar{1}\ 0]$ and $Al_2O_3$ $[1\ 1\ \bar{2}\ 0]$ planes.[33] The in-plane spacings of TiN $(1\ 1\ \bar{2})$ and $Al_2O_3$ $(2\ \bar{1}\ \bar{1}\ 0)$ are 1.71 and 2.38 Å, respectively, giving a 4:3 coincidence of 6.85 and 7.14 Å with a mismatch of 4%. This close coincidence is consistent with substrate-templated domain-matching epitaxy at the TiN–sapphire interface.

Crystal coherent length (CCL) is a crucial parameter for quantifying the degree of structural order of a solid-state thin film. The CCL represents the distance over which the crystal lattice maintains a well-defined periodic structure along a given crystallographic direction, determined by the width of Bragg peaks using the Scherrer equation in reciprocal space[37], $D_{hkl} = 2\pi K/\Delta q_{hkl}$, where $\Delta q_{hkl}$ is the full width at half maximum of the corresponding Bragg peak, $q_{hkl}$ is the scattering vector and, and $K = 0.9$ is the Scherrer constant. The CCL value is estimated to be 14.9, 15.8 and 19.5 nm for (200), (111) and (220) peaks respectively. Given the twinned structure of our TiN film, the CCL can be considered as an upper limit for the average crystal domain size between twinned boundaries. Conservatively, due to the presence of twinned domains in the TiN film, we take the lower limit of ~15 nm.

Neutron reflectometry provides a measure of film thickness and nuclear scattering length density with a high sensitivity to hydrogen content in the active layer. Figure 1d shows neutron reflectivity fitted with reflectivity simulations. The fit result shows a thickness of 42.9 nm and a scattering length density of $3.053\times10^{-6}$ Å$^{-2}$. The deviation from the theoretical value ($3.204\times10^{-6}$ Å$^{-2}$), derived from the measured lattice parameter (4.197 Å), corresponds to a hydrogen concentration of 3.7% in the active layer, suggesting a stoichiometric TiN and no evidence for substantial hydrogen uptake during the ALD process (see Supplementary Section 3)[38,39].

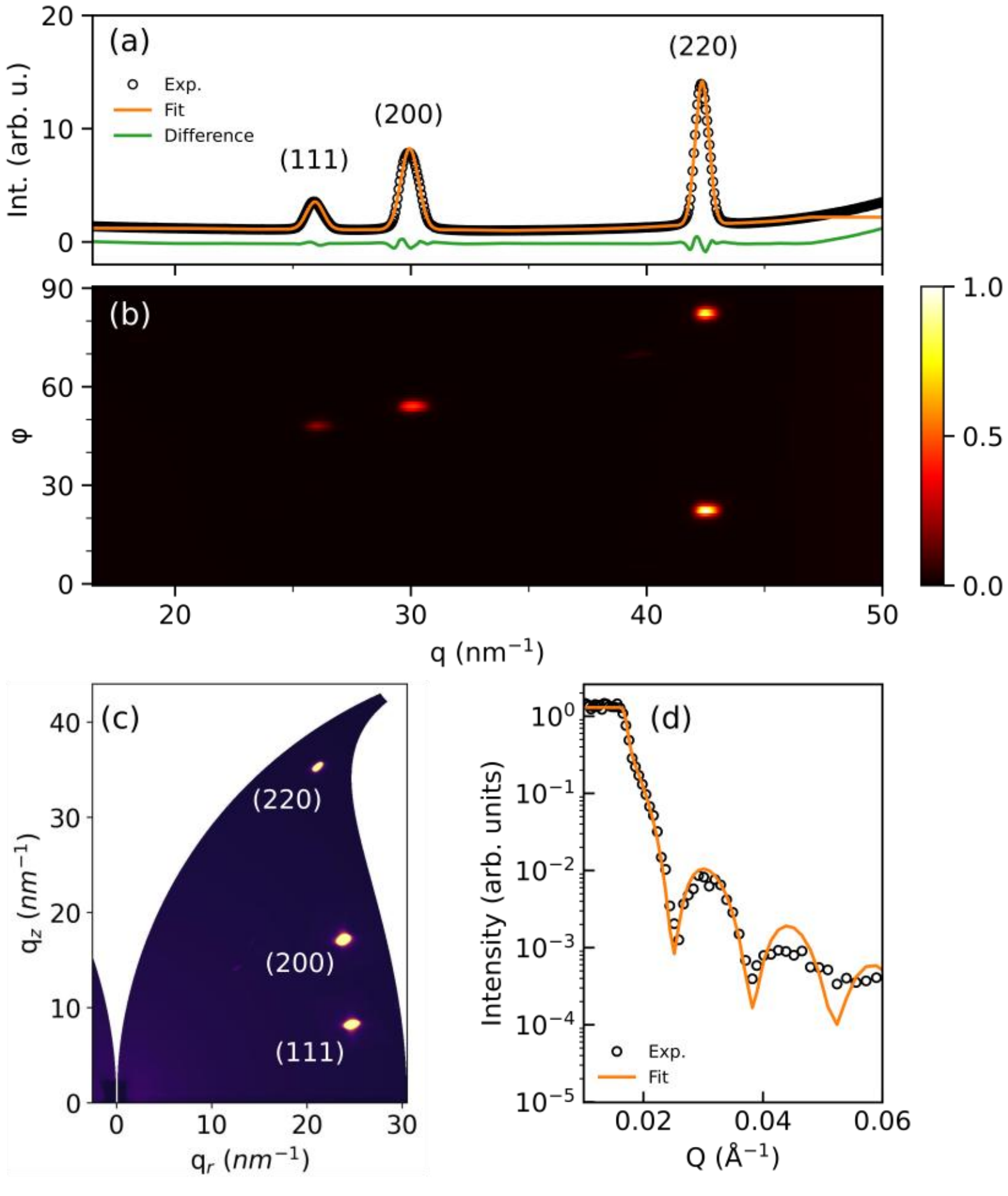

Figure 1. Structural characteristics of TiN thin films. (a) 1D profile as a function of total reciprocal lattice vector, $q$, integrated over GIWAXS 2D images. Bragg peaks were observed at $q$ = 25.9, 29.9, 42.3 nm$^{-1}$. The experimental data are fit using the Le Bail method to obtain the unit cell symmetry and lattice parameters. (b) Evolution of 1D integrated profiles as a function of the sample rotation angle, φ, recorded through a φ rotation around the film normal. Inset: a schematic of φ angle relative to the sample. (c) 2D GIWAXS pattern obtained at an incidence angle of 0.28°. The pattern is generated from multiple 2D images at different φ. (d) Neutron reflectivity as a function of $q$ fitted with simulation.

Superconducting properties of TiN films grown by HPA-ALD are shown in Figure 2. A superconducting transition temperature $T_c$ is observed at 2.2 K, shown in temperature-dependent sheet resistance $R_{sq}$ (Figure 2a), and the temperature dependence of $R_{sq}$ follows Bloch-Grüneisen behaviour (see the inset of Figure 2a). Below 50 K, $R_{sq}$ saturates at a residual resistance $R$ = 22.7 Ohm/sq corresponding to a residual resistivity ratio of 1.11, comparable to that of superconducting TiN thin films grown by other methods.[40] The temperature-dependent measurement allows us to extract the kinetic inductance $L_K$, defined as $L_K = \hbar R_s/\pi\Delta_0$, where $R_s$ is the resistance measured just above $T_c$, and the superconducting gap $\Delta_0 = 1.76 k_B T_C$ is obtained from the Bardeen-Cooper-Schrieffer theory. Our TiN films obtain a $L_K$ = 15 pH/sq, which is higher than that of standard ALD-grown TiN films of similar film thickness[28,41]. The upper critical field and superconducting coherence length are intrinsic physical properties of a superconductor. Figure 2b shows the evolution of $R_{sq}$ as a function of out-of-plane magnetic field $B_\perp$ at temperatures ranging from 0.55 to 2.1 K. A transition of superconducting–mixed–normal state is observed, indicating a type II superconductor and enabling extraction of the upper critical fields $H_{c2}$ as a function of temperature. $H_{c2}$ is defined by a linear extrapolation between the normal state and mixed state. The temperature dependence of $H_{c2}$ is approximately linear (see Figure 2c), particularly close to $T_c$, consistent with Ginzburg-Landau (GL) theory. We then determined the GL coherence length[42] ($\xi_0$) using, $(\xi_0)^2 = \Phi_0/2\pi T_c H_{c2}(0)$, $H_{c2}(0) = 0.69 T_c |dH_{c2}(T)/dT|_{T\approx T_c}$, where $\Phi_0$ is the magnetic flux quantum, resulting $\xi_0$ = 14.6 nm. Using measured resistivity, we obtained the electron diffusion constant $D$ = 1.20 cm$^2$s$^{-1}$, calculated using the Einstein relation, $\sigma = e^2 N(E_F) D$, where $N = 5.4\times10^{22}$ eV$^{-1}$cm$^{-3}$ is carrier concentration, $\rho = 1/\sigma$ = 96.5 μΩ·cm is conductivity of the film. In dirty-limit superconductors, the superconducting coherence length can be calculated by $\xi = \sqrt{\hbar D/\Delta_0}$, which results in $\xi$ = 15.4 nm. This shows a self-consistent result compared with the experimentally determined value. Using the Drude model, $l = 3D/v_F$, where $v_F = 7\times10^5$ ms$^{-1}$ is the Fermi velocity, we obtained the electron free mean path $l = 0.5$ nm, close to the size of unit cell ($a = 0.4197$ nm), suggesting that the film approaches the Ioffe-Regel limit, indicating a strongly disordered metal. This indicates that structural order and electronic disorder are decoupled in our TiN, with GIWAXS study showing long-range order and electrical transport showing atomic-scale scattering. We note that the TiN obtains the experimentally determined $\xi_0$ comparable to the CCL obtained from GIWAXS measurements, suggesting a possible correlation of the coherence length of Cooper pair to fine structural features of epitaxy films, such as twinned boundaries[43]. Further investigation will be required to link the coherence length to a specific structural feature.

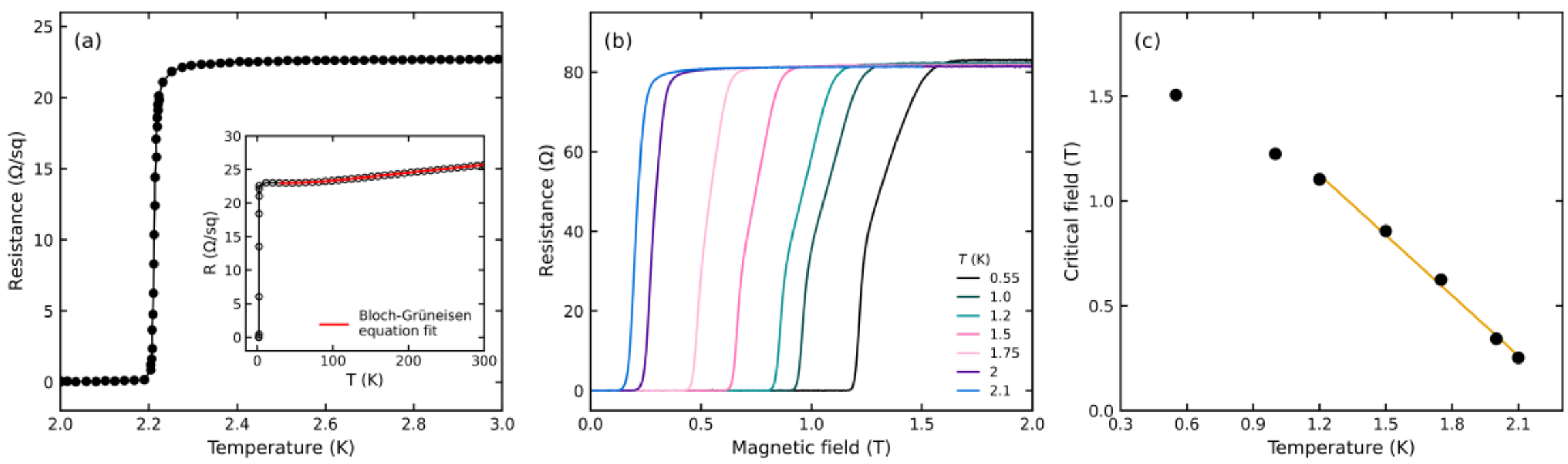

Figure 2. Superconducting properties of TiN films. (a) Temperature dependence of sheet resistance $R_{sq}$ of a TiN Hall bar sample. We obtained a critical temperature $T_c$ of superconductivity at 2.2 K. Inset: temperature-dependent $R_{sq}$. We obtained $n = 4.8$ and an estimated Debye temperature $\theta_D \sim 470$ K using Bloch-Grüneisen formula. (b,c) Magnetotransport properties of evolution of resistance as a function of out-of-plane magnetic field with temperature ranging from 0.55 to 2.1 K (b), and extracted upper critical field $H_{c2}$ as a function of temperature (c).

Chemical composition of superconducting TiN was investigated using synchrotron-based soft X-ray spectroscopy. Figures 3a–f show Ti 2p, N 1s, and O 1s core level spectra of as-introduced (Figures 3a–c) and sputtered sample (Figures 3d–f). Detailed fit parameters are summarised in Supplementary Information. The as-introduced sample exhibits doublets associated with oxidic, Ti–O, and oxynitride, Ti–O–N, species, supported by related chemical shifts in the observed in Ti 2p, O 1s and N 1s core levels. The gentle sputtering reduces the peak intensity of oxide and oxynitrides relative to the Ti–N bond, indicating the removal of surface oxide. We observed contribution of surface core level shifts induced by hydrogen adsorption on TiN surface[44–46] resulting multiple peaks (filled pentagons) exclusively in the sputtered sample, which can be attributed to highly surface-sensitive measurement due to the fixed kinetic energy of photoelectrons of 70 eV (corresponding to an inelastic mean free path $\lambda_{IMFP}$ = 4.6 Å). Accordingly, residual surface oxide remains detectable in sputtered TiN due to the combination of incomplete removal of surface oxide and the shallow measurement depth ($3\times\lambda_{IMFP}$ = 13.8 Å). We rule out band bending effects due to the metallic nature of TiN. Figure 3g shows Ti L-edge spectra recorded in total electron yield (TEY) mode before and after Ar sputtering. The Ar sputtering results in an overall shift of $L_{2,3}$-edge transitions to lower photon energies, and significantly reduces the peaks at the lower-energy side of $L_{2,3}$-edge transitions (filled black circles) which indicates Ti–O bonds[47]. Since an oxidised environment is expected to have a higher energy onset than un-oxidised one, the shifting indicates the manifestation of un-oxidised TiN underlaying surface oxide. Given the analytic depth of TEY spectra (~10 nm), the absence of Ti–O bonds in sputtered TiN suggest that oxygen is confined to a surface layer of order 1 nm with no evidence for oxide in the near-surface bulk within the TEY probe depth.

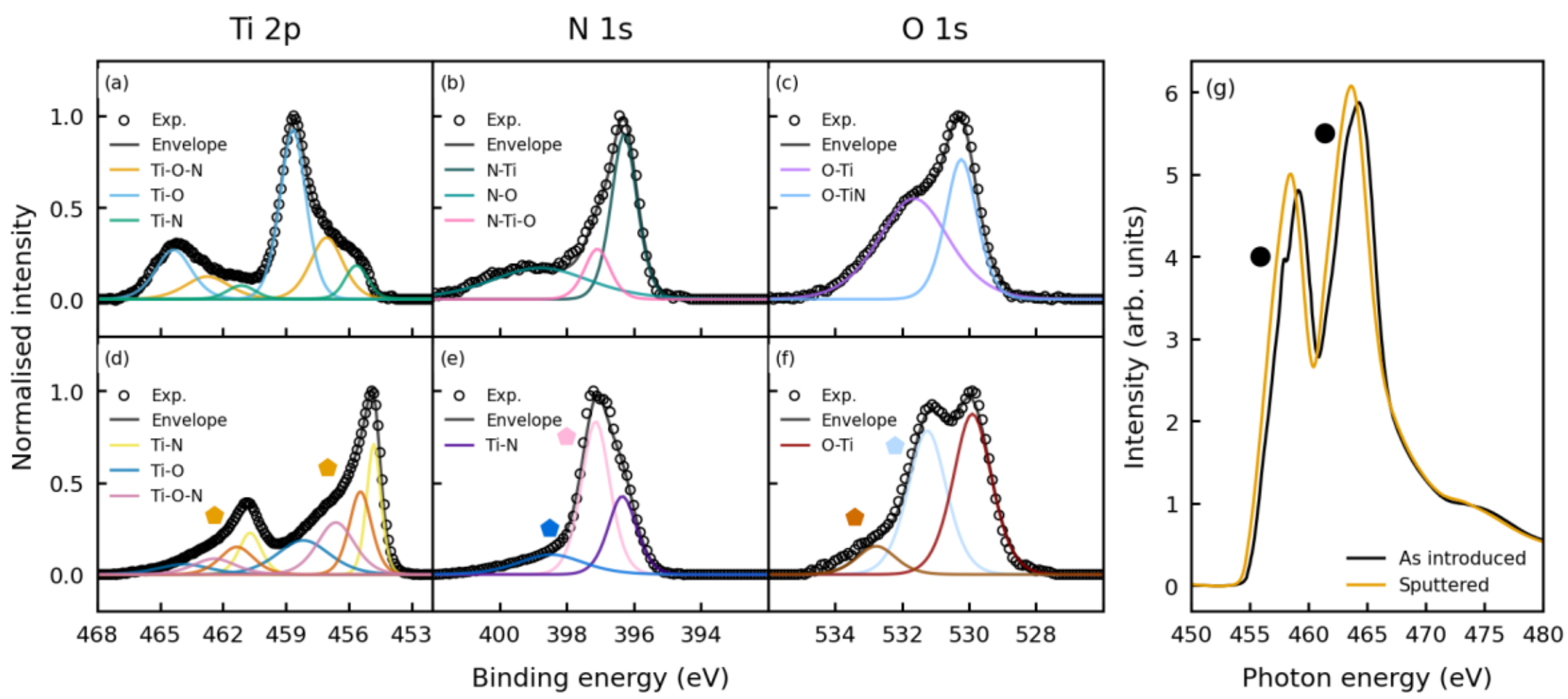

Figure 3. Compositional analysis of HPA-ALD grown TiN. (a–f) High-resolution XPS spectra of Ti 2p, N 1s, and O 1s core levels for the as-introduced (a–c) and sputtered sample (d–f). Black circles and solid lines represent the experimental data and the fitted envelopes respectively. All the spectra were obtained with a fixed kinetic energy of 70 eV. The fit

parameters and survey scan for each measurement can be found in Supplementary Information. (g) Ti $L_{2,3}$-edge XAS spectra collected in total electron yield mode for the as-introduced and sputtered samples.

In conclusion, we have successfully demonstrated HPA-ALD as a feasible tool to grow high-quality superconducting TiN thin films. Complementary synchrotron-based spectroscopy techniques confirm the semi-coherent epitaxy templated by sapphire substrate, exhibiting minimal oxygen and hydrogen impurities throughout the film and minimal structural disorder specifically at the TiN-sapphire interface. This places the HPA-ALD as a promising method for preparing high-quality superconducting thin films in manufacturing wafer-scale superconducting devices. Electrical transport measurements show favourable superconducting properties compared to reported studies, suggesting a more versatile option for harnessing the kinetic inductance in superconducting quantum device applications. The coincidence between CCL and superconducting coherence length could be intriguing for further investigation. Our findings highlight HPA-ALD as a promising epitaxial method for superconducting quantum research and applications.

**Methods**

HPA-ALD process for TiN thin films

TiN thin films were deposited via a cyclic HPA-ALD process. In each ALD cycle, tetrakis(dimethylamino)titanium (TDMAT) was first introduced as the titanium precursor, followed by an Ar purge. Subsequently, $H_2$ plasma was applied at a power of 300W for a duration of 40s to improve surface chemistry. After the second Ar purge step, $N_2$ plasma was introduced under the identical plasma power and duration. A final argon purge was then performed to eliminate remaining reactive species, thereby completing one full ALD cycle.

TiN sample preparation for electrical transport measurements

Superconducting TiN thin films were patterned into a Hall bar geometry using standard optical lithography and dry etching. Dry etching process was performed using reactive ion etching (RIE) at room temperature in a mixed Ar/$SF_6$ gas (Ar: $SF_6$ = 15: 9). The chamber pressure was maintained at 10 mTorr and the RF power set to 90 W. For DC transport measurements, electrical characterisation was performed in a Physical Property Measurement System in a helium environment at a pressure of 20 Torr. The channel conductance was measured using a Stanford Research Systems lock-in amplifier. For magnetotransport measurements, electrical characterisation was in a dilution refrigerator at a base pressor of $10^{-6}$ Torr. The measurements were performed using a source current below the critical current to avoid possible Joule heating

Synchrotron-Based Grazing-Incidence Wide-Angle Scattering

Room-temperature GIWAXS data of TiN films were recorded at the NCD-SWEET beamline at ALBA synchrotron (Cerdanyola del Vallès, Spain) using a monochromatic X-ray beam ($\lambda$ = 1.34 Å) with a size of 150 × 25 $\mu m^2$ (H × V). The beam energy was selected using a Si (111) channel cut monochromator, which was collimated using an array of Be collimating lenses. The scattered signal was recorded using a model LX255-HS area detector (Rayonix, LLC, Evanston, IL) placed 229.4 mm from the sample position. Detector tilts and the sample–detector distance were calculated using $Cr_2O_3$ as the calibration standard. GIWAXS frames were recorded at incident angles of ($\alpha_c$ = 0.28°), which is above the critical angle of the film, to acquire bulk-sensitive scattering regime of the TiN film. Sample rotations around the normal direction were realised using a stepper motor and a continuous $N_2$ flow over the sample was employed during the measurement to ensure the sample remained in a chemically inert atmosphere during irradiation. Collected 2D images were azimuthally integrated using PyFAI and processed using a custom python routine.

Synchrotron-based X-ray photoelectron spectroscopy (XPS)

XPS data were recorded at the soft x-ray beamline (Australian synchrotron in Melbourne, Australia) under ultra-high vacuum (UHV) Prevac Endstation (base pressure 1 × $10^{-10}$ mbar). Samples were cleaned in a dedicated prep chamber connected to the analysis chamber. Surface oxide was removed by Ar ion sputtering with energy of 500 eV, and confirmed by XPS and Near Edge X-ray Absorption Spectroscopy (NEXAFS) spectrum. All XPS data were calibrated to Au $4f_{7/2}$ peak with binding energy of 83.4 eV. For NEXAFS spectra, the spectra were normalised by incident X-ray intensity and set the pre-absorption signal as zero.

ACKNOWLEDGEMENT

This research was supported by the Australian Government through the Australian Research Council's Linkage Projects funding scheme (project LP210200636). J. A. S. acknowledges financial support from the Australian Research Council (DE230100173, DP260103534). This research was supported by a UQ Amplify Fellowship from The University of Queensland. We acknowledge travel funding provided by the International Synchrotron Access Program (ISAP) managed by the Australian Synchrotron, part of ANSTO, and funded by the Australian Government. The authors thank the staff of the BL11 NCD-SWEET beamline at ALBA Synchrotron for their assistance in recording the GIWAXS data. Y. C. H. and M. J. C. acknowledge the financial support in part by the National Science and Technology Council (NSTC 113-2221-E-002-091-MY3, 115-2640-E-002-018, and 114-2622-8-002-016), Taiwan. This work used the Queensland node of the NCRIS-enabled Australian National Fabrication Facility (ANFF).

AUTHOR CONTRIBUTION

Y. H. C. conceived, designed, executed throughout this project. Y. C. H and M. J. C. contributed to material growth. Y. H. C., Z. D., E. S., D. S., D. C., P. J., and J. A. S. performed neutron reflectometry and synchrotron experiments, and data analysis. Y. H. C., K. X., M. B., M. S. F. and A. F. contributed to device fabrication and electrical transport measurements. The original draft was written by Y. H. C. The finalised manuscript was written through contributions of all authors. All authors have given approval to the final version of the manuscript.

NOTES

The authors declare no competing interest.